\documentclass[preprint2]{aastex}

\shorttitle{Conjugate OH Toward PKS 1413+135}
\shortauthors{Darling}

\begin{document}


\title{A Laboratory for 
Constraining Cosmic
Evolution of the Fine Structure Constant:
Conjugate 18 cm OH Lines Toward PKS 1413+135 at $z=0.2467$
}


\author{Jeremy Darling}
\affil{Carnegie Observatories, 813 Santa Barbara Street, 
    Pasadena, CA 91101}
\email{darling@ociw.edu}



\begin{abstract}
We report the detection of the satellite 18 cm OH lines at 1612 and 1720 MHz
in the $z=0.2467$ molecular absorption system toward the radio source PKS
1413+135.  The two OH lines are conjugate; the 1612 MHz line 
is seen in absorption while the 1720 MHz line is seen in weak maser emission 
of equal, but negative, optical depth.  
We do not detect the main 18 cm OH lines at 1667 and 1665 MHz
down to 1.1 mJy rms in 4.0 km s$^{-1}$ channels.
The detected and undetected 18 cm OH lines support a scenario of 
radiatively pumped stimulated 
absorption and emission with pumping dominated by the intraladder 119 
$\mu$m line of OH, 
suggesting a column density $N(OH) \simeq 10^{15}$--$10^{16}$ cm$^{-2}$.
Combined with simultaneous \ion{H}{1} 21 cm observations and published CO 
data, we apply the OH redshifts to
measurements of cosmic evolution of the fine structure constant 
($\alpha\equiv e^2/\hbar c$).
We obtain highly significant ($\sim25\sigma$)
velocity offsets between the OH and \ion{H}{1} lines and the OH and CO lines,
but measurements of $\alpha$-independent systematics
demonstrate that the observed velocity differences are
entirely attributable to physical velocity offsets between species rather
than a change in $\alpha$.
The OH alone, in which conjugate line profiles
guarantee that both lines originate in the same molecular gas,
provides a weak constraint
of $\Delta\alpha/\alpha_\circ = (+0.51 \pm 1.26) \times 10^{-5}$ at $z=0.2467$.
Higher frequency OH line detections can provide
a larger lever arm on $\Delta\alpha$ and can increase precision by an order of
magnitude.  
The OH molecule can thus provide precise measurements 
of the cosmic evolution 
of $\alpha$ that include quantitative constraints on systematic errors.  
Application of this technique is limited only by the detectability of
$|\tau|\sim0.01$ OH lines toward radio continuum sources and may
be possible to $z\sim5$.  
\end{abstract}





\section{Introduction}

Recent work to investigate the cosmic evolution of the fine structure 
constant ($\alpha\equiv e^2/\hbar c$) suggests a decrement in the past:
$\Delta\alpha/\alpha_\circ = (-0.574\pm0.102)\times10^{-5}$ for 
$0.2\leq z\leq 3.7$, where $\Delta\alpha\equiv\alpha(t)-\alpha_\circ$
\citep{mur03}.  This result applies the ``many-multiplet''
method to optical absorption lines in the spectra of high redshift quasars;
$\alpha$ parameterizes the strength of electromagnetic interactions and
a changing $\alpha$ will cause spectral line shifts from present-day values.
\citet{sri04} apply the same method to a new sample of \ion{Mg}{2} absorption 
systems and find no change in $\alpha$ over $0.4\leq z\leq 2.3$:  
$\Delta\alpha/\alpha_\circ = (-0.06\pm0.06)\times10^{-5}$.  
Other methods, including terrestrial measurements from the 
Oklo natural fission reactor and meteorites provide mixed results: 
from Oklo, \citet{fuj00} obtain 
$\Delta\alpha/\alpha_\circ=(0.4\pm1.6)\times10^{-8}$ while
\citet{lam04} obtain $\Delta\alpha/\alpha_\circ\geq4.5\times10^{-8}$ 
with 6$\sigma$ confidence, both at $z\simeq0.16$ (2 Gyr); and 
\citet{olive04} obtain $\Delta\alpha/\alpha = (0.08\pm0.08)\times10^{-5}$ 
at $z\simeq0.45$ from meteorites.
Other astrophysical constraints obtain results consistent with no change
in the fine structure constant:  
\ion{O}{3} emission line shifts in quasars with $0.16<z<0.80$ from the 
Sloan Digitized Sky Survey obtain $\Delta\alpha/\alpha=(7\pm14)\times10^{-5}$
\citep{bah04}; and comparisons of
\ion{H}{1} 21 cm (1420 MHz) line redshifts ($\nu\propto\alpha^4$)
to molecular rotation transitions ($\nu\propto\alpha^2$) in molecular 
absorption systems obtain $\Delta\alpha/\alpha=(-0.10\pm0.22)\times10^{-5}$ 
at $z=0.2467$ and $\Delta\alpha/\alpha=(-0.08\pm0.27)\times10^{-5}$ 
at $z=0.6847$ \citep{mur01}.

Radio observations of molecular and \ion{H}{1} absorption lines show 
much promise for exceptionally precise measurements of the value of $\alpha$
across a large span of cosmic time.  These lines are generally narrow and 
simple.  Complex molecules and molecules with unbalanced electron angular
momentum, like OH and CH, offer multiple transitions with differing dependence
on $\alpha$, providing measurements of $\Delta\alpha$ from a {\it single 
species} including quantitative estimates of systematic velocity offsets 
between species \citep{dar03}.  These same molecules can also provide 
simultaneous constraints on the cosmic variation of $\alpha$, $g_p$, and
$m_e/m_p$ \citep{che03,kan04}.

Further, OH can form a natural maser across a wide range of physical settings, 
and observed
ratios of the four ground state 18 cm OH lines, at 1612, 1665, 1667, and 1720 
MHz, provide clues to the physical
conditions associated with the masing.
A special case
occurs when both emission and absorption are stimulated in pairs of 
{\it conjugate} OH lines, lines that are identical except for the sign
of the optical depth.  Conjugate lines are a direct 
result of quantum selection rules and an optically thick far-infrared
transition from a single dominant excited rotation state 
\citep{elit92,vanlang95}.  Conjugate lines can be added to obtain a perfect
null, indicating that the lines are produced in the same molecular gas complex.
Hence, conjugate lines provide a valuable laboratory for precision 
measurements of fundamental constants:  lines are guaranteed to reside in the
same physical location at the same physical velocity, so any relative
line shift in a conjugate pair indicates a change in physical constants
rather than systematic effects.

The primary difficulty with the molecular absorption line approach to 
changing constants is the dearth of molecular absorption systems.  To our
knowledge, only six cm- and mm-detected molecular absorbers at
$z\gtrsim0.1$ are known, only four of these have $z>0.2$, and the highest 
redshift system is PKS $1830-211$ at $z=0.8858$
\citep{wik96,che99}.
Numerous groups are currently searching for additional
high redshift mm molecular 
absorption systems (e.g. \citet{mur03a}).

This letter presents new observations of the OH 18 cm and HI 21 cm lines 
toward PKS 1413+135 at $z=0.2467$ that reveal conjugate OH satellite lines
(\S \ref{sec:obs},\ref{sec:results}).
The conjugate lines are guaranteed to originate in the same gas complex (\S
\ref{sec:model}) and provide a new laboratory for a measurement of the
fine structure constant at $z=0.2467$ (\S \ref{sec:alpha}).  
The OH lines also provide quantitative
estimates of the systematic velocity offsets between species (such as OH, 
HI, and CO), demonstrating that seemingly significant nonzero 
$\Delta\alpha/\alpha_\circ$ measurements are in fact consistent with zero once
systematic effects have been removed.  

\section{Observations}\label{sec:obs}


We observed the molecular absorption system at $z=0.2467$ toward 
the radio source PKS 1413+135 at the NRAO Green Bank Telescope\footnote{The 
National Radio Astronomy Observatory is a facility of the 
National Science Foundation operated under cooperative agreement by 
Associated Universities, Inc.} on December 14, 17, and 19, 2003, simultaneously
observing the four $^2\Pi_{3/2}\ J=3/2$ 
OH lines at 1612.23101, 1665.40184, 1667.35903, and
1720.52998 MHz and the \ion{H}{1} line at 1420.405751786 MHz, 
each appropriately
redshifted and tracked in a barycentric reference frame.  We observed four
12.5 MHz bandpasses centered on the 1420, 1612, 1667, and 1720 MHz lines
in two linear polarizations in a 5 minute position-switched mode with a 
winking calibration diode and data recorded every 0.5 seconds. Bandpasses were
divided into 8192 channels, but the nature of the autocorrelation
spectrometer reduced the effective spectral resolution to 3.05 kHz or 0.80
km s$^{-1}$ for \ion{H}{1} at $z=0.2467$.  
Fast sampling was intended to facilitate radio
frequency interference (RFI) excision, but the observed bands were generally
clean in the vicinity of the observed lines.  The OH 1720 MHz line did show
some nearby RFI, but careful inspection and flagging eliminated any possible
disruption or contamination of the line profile.  

Records were individually calibrated and bandpasses flattened using the 
winking calibration diode and the corresponding off-source records.
Records and polarizations were subsequently averaged
to obtain a final spectrum.  
Spectra were Hanning smoothed, and a linear or quadratic baseline was 
fit and subtracted from a narrow spectral window centered on each line 
(including the 1665 MHz line).
The total on-source integration time was 90 minutes for the OH lines and
120 minutes for the \ion{H}{1} line.  We reached a typical
rms noise of 1.8 mJy in 3.05 kHz channels.  
All data reduction was performed in 
AIPS++\footnote{The AIPS++ (Astronomical Information Processing System) 
is freely available for use under 
the Gnu Public License. Further information may be obtained from 
http://aips2.nrao.edu}.


Our single dish flux density measurements of PKS 1413+135 are consistent
with previous observations.  At 1140 MHz, the observed flux density is 
1.367 Jy, consistent with the \citet{car92} measurement of $1.25\pm0.15$ Jy.
At 1380 MHz, we obtain 1.085 Jy and \citet{con98} measured $1.092\pm33$ Jy.
At 1293 and 1337 MHz, we obtain 1.162 and 1.192 Jy respectively.  Flux 
calibration of these data seems to be reliable, but can be systematically 
affected by in- and out-of-band RFI, especially in the 1140 MHz band which 
contains abundant strong RFI.  Systematic flux calibration errors in these
data appear to be well below $10\%$.  




\section{Results} \label{sec:results}

\begin{figure*}
\epsscale{1.80}
\plotone{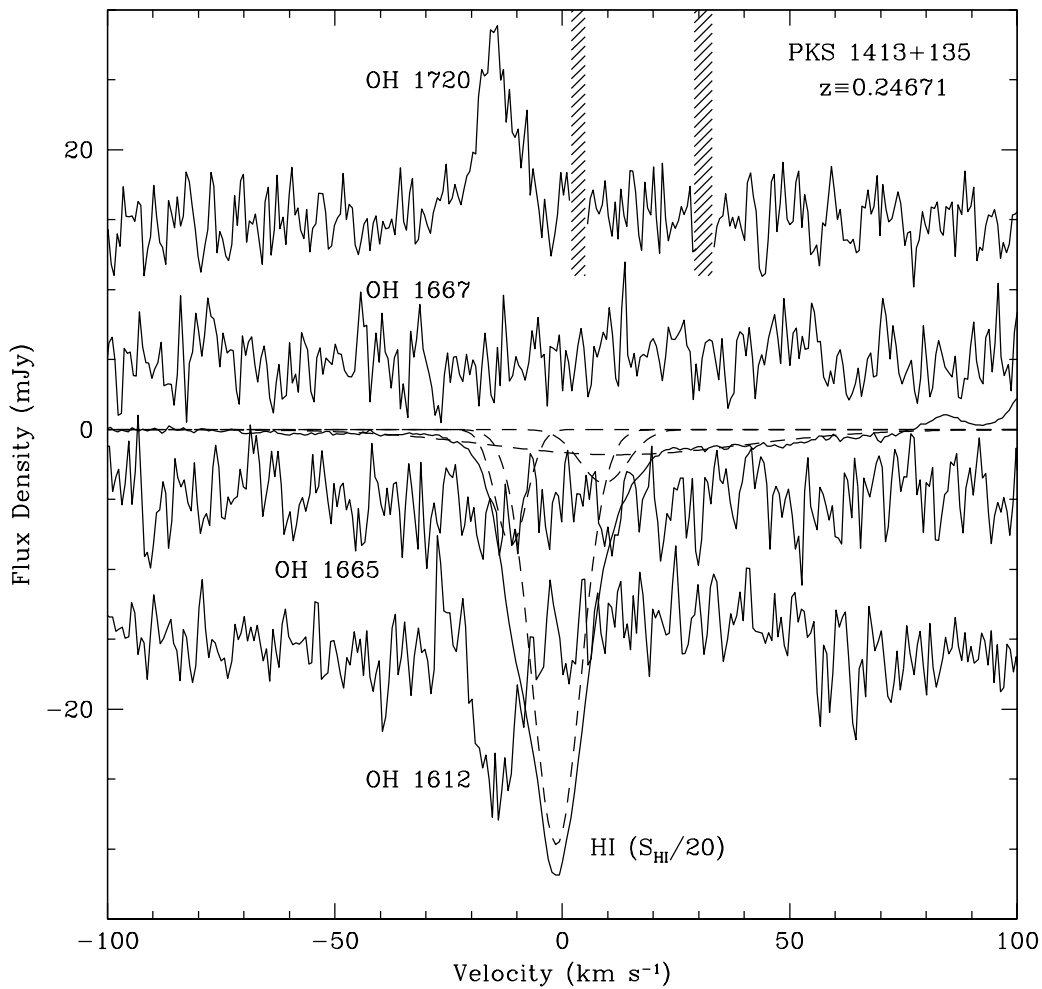}
\caption{OH and \ion{H}{1} lines at $z=0.24671$ toward PKS 1413+135.  All lines
were observed in a barycentric frame, zero velocity refers to a redshift
of $z=0.24671$, and the velocity scale is in the rest frame of the
absorption system.  A linear or quadratic fit to the spectral baseline 
has been subtracted from each spectrum.
The \ion{H}{1} spectrum has been scaled down by a factor of
20 and the OH spectra have been offset from zero by $-15$, $-5$, $5$, and
$15$ mJy for the 1612, 1665, 1667, and 1720 MHz lines, respectively.
The dashed lines show the 4 gaussian components fit to the \ion{H}{1}
profile.  The positive features in the \ion{H}{1} spectrum are RFI, and 
the broad \ion{H}{1} component (\ion{H}{1} 4) may be an artifact of 
this interference.
Shaded regions in the 1720 MHz spectrum at 31 and 4 km s$^{-1}$ are 
regions where
RFI was excised and no on-sky data exists.
\label{fig1}}
\end{figure*}

Of the five lines observed toward PKS 1413+135, only three were detected:
the 1612 and 1720 MHz lines of OH and the 21 cm line of \ion{H}{1} 
(Figure \ref{fig1}).  
The \ion{H}{1} and 1612 MHz lines were detected in absorption.  The
1720 MHz line has a profile similar to the 1612 MHz line profile modulo 
the spectral noise, but with a negative optical depth.
The 1667 MHz line is not detected, contrary to the detection by 
\citet{kan02}.  We obtain an rms noise value of 1.1 mJy in boxcar smoothed
spectra with 4.0 km s$^{-1}$ resolution similar to the resolution obtained
by \citet{kan02}, who detect the 1667 MHz line in absorption at 7.9 mJy.  
The 1665 MHz line is not detected at a similar limit to the 1667 MHz line
(Table \ref{table:fits}).  
Gaussian fits to the detected lines are listed in Table \ref{table:fits}; the
OH lines are well fit by single gaussian profiles, but the \ion{H}{1}
profile requires 4 components.  The broad \ion{H}{1} 4 component may 
be an artifact associated with the RFI above $\sim80$ km s$^{-1}$.
To within the quoted uncertainties in 
the gaussian fits, the two OH line profiles are indistinguishable in 
redshift, width, and $|\tau|$.  

Although there is evidence for a nonzero spectral index across the
observed bands, optical depths are computed assuming a constant flux 
density of 1.2 Jy at 1100--1400 
MHz for PKS 1413+135.  This assumption is broadly consistent with 
the observations at the $10\%$ uncertainty level and simplifies comparisons
with previous observations.  The core-jet radio continuum of 
PKS 1413+135 is extended on scales of 60 
milliarcseconds (215 pc) and the \ion{H}{1} absorption is 
certainly associated with the jet, although it may extend across the core
as well \citep{car00,perl02}.  Optical depths are thus likely 
to be lower bounds because the continuum covering factor is unknown.  

Despite the uniform assumptions about the source continuum level, we
observe a significantly deeper \ion{H}{1} absorption 
($\tau_c = 0.757\pm0.002$, statistical error only)
than \citet{car92} who obtained a peak optical depth of $0.34\pm0.04$.  
Given the incomplete covering of the continuum source,
our result is compatible with $\tau=0.89$ obtained from
VLBI observations by \citet{car00}.  
Note that the quoted errors in the \ion{H}{1} profile fits are very 
conservative and take into account the blended line profile; 
fitting with fewer components would give a lower
uncertainty in the line position but a larger residual to the fit.
Note also that a single component gaussian fit to the \ion{H}{1} 
asymmetric profile produces an offset from the peak optical depth.


%
%

%
%



\section{Conjugate OH Lines} \label{sec:model}

The optical depths in the 18 cm OH satellite lines are of order $\pm0.01$ 
while the main lines are suppressed by an order of magnitude, below 
$\tau \sim 0.001$.  The observed conjugate satellite lines and the suppressed
main lines are consistent with 
a scenario of radiatively pumped stimulated 
absorption and emission with pumping dominated by 
optically thick radiative decay from the $^2\Pi_{3/2}\ J=5/2$ rotation state
119 $\mu$m above the OH ground state $^2\Pi_{3/2}\ J=3/2$.  As explained by
\citet{elit92} and \citet{vanlang95} the 18 cm transitions compete for the 
same IR pumping photons.
Since the population of the ground states is governed by the selection rule
$\Delta F = \pm1,0$ (and a change in parity) and if the radiative 
decay to the ground state is 
optically thick, then the $F=2,3$ states of the $J=5/2$ level overpopulate
the $F=2$ states compared to the $F=1$ states of the ground $J=3/2$ level
($\Delta F=\pm2$ is forbidden).
The main OH lines follow the rule $\Delta F=0$ while the satellite lines 
have $\Delta F=\pm1$.  Since 1720 MHz emission follows $F=2^+\rightarrow1^-$,
it is seen in emission.  The 1612 MHz line appears as stimulated absorption
via $F=2^-\rightarrow1^+$.  The conjugate behavior of the satellite lines
maintains equal populations in the two $F=1$ and equal populations in 
the two $F=2$ states, 
suppressing nonthermal $\Delta F=0$ main line (1665 and 1667 MHz) 
emission or absorption.

Extragalactic conjugate OH satellite lines have been observed toward 
NGC 4945 \citep{whi75},
Cen A \citep{vanlang95}, M82 \citep{sea97}, and NGC 253 \citep{fra98}.
In some cases, like NGC 253, the OH satellite lines are conjugate over a 
wide range of conditions and show cross-over from absorption to emission 
(and vice-versa).  The 1612 MHz line changes to emission when the radiative
pumping becomes dominated by the cross-ladder 79 $\mu$m transition from 
the $^2\Pi_{1/2}\ J=1/2$ state \citep{vanlang95}.

The scenario proposed for PKS 1413+135
is valid for an optically thick 119 $\mu$m intraladder
transition but requires an optically thin 79 $\mu$m cross-ladder transition.
Since the intraladder and cross-ladder Einstein coefficients differ
by nearly an order of magnitude, there is an order of magnitude range 
in the OH column density 
$10^{14}$ cm$^{-2}$ km$^{-1}$ s $ \lesssim N_{OH}/\Delta V \lesssim 10^{15}$ 
cm$^{-2}$ km$^{-1}$ s for conjugate behavior with 1720 MHz in emission
\citep{vanlang95}.
For $\Delta V \approx 10$ km s$^{-1}$, $N_{OH} \approx 10^{15}$--$10^{16}$ 
cm$^{-2}$.  Assuming an OH abundance of $N_{OH} \sim 10^{-7}\ N_{H_2}$, 
we obtain an order of magnitude estimate of the molecular hydrogen column 
of $N_{H_2} \approx 10^{22}$--$10^{23}$ cm$^{-2}$.  
This is significantly different from CO estimates by \citet{wik97} of
$N_{H_2} \geq 4\times10^{20}$ cm$^{-2}$ which they stress may be a severe
underestimate for a dense clumpy medium.  
The CO may also be in a different physical region than the OH 
(\S \ref{sec:alpha}).
Note that the usual direct determination of the OH column from the integrated
optical depth \citep{lis96} is inappropriate for this case of stimulated
transitions.


\section{Measuring the Fine Structure Constant} \label{sec:alpha}

OH lines can provide precise constraints on cosmic evolution of the physical
constants $\alpha$, the proton g-factor, and the ratio of electron
to proton mass \citep{dar03,che03,kan04}.  
The following treatment assumes that $\alpha$ evolution provides the
dominant contribution to any changes in line rest frequencies.  Since
only the satellite 18 cm OH lines have been detected in PKS 1413+135, 
we provide supplemental equations to \citet{dar03} which focused on the 
main OH lines.  

The 18 cm OH lines can be decomposed into a $\Lambda$-doubled term 
which depends weakly on $\alpha$ and a hyperfine term which has the 
usual strong $\alpha^4$ dependence \citep{dar03}.  From these, sums and
differences of lines can form pure $\Lambda$-doubled and pure hyperfine
quantities:
\begin{eqnarray}
  \Sigma\nu &\equiv& \nu_{1720}+\nu_{1612} = 2\Lambda\alpha^{0.4}\\
  \Delta\nu &\equiv& \nu_{1720}-\nu_{1612} = 2(\Delta^+ + \Delta^-) \alpha^4
\end{eqnarray}
where $\Delta^+=9.720353(25)\times10^9$ MHz and 
$\Delta^-=9.375256(25)\times10^9$ MHz set the strength of 
hyperfine splitting, $\Lambda=11926.36309(51)$ MHz sets
the strength of the $\Lambda$-doubling, and the exponents on $\alpha$ are
accurate to $\leq5\%$ \citep{dar03}.  Redshift differences
between pairs of OH lines can be tied directly to changes in $\alpha$, 
mitigating
systematic effects by comparing highly correlated lines from a single species:
\begin{equation}
	{z_{1612} - z_{1720} \over 1+z_{1720}}
		\simeq 1.8 \left(\Delta\alpha\over\alpha_\circ\right)\,
		\left(\Sigma\nu\ \Delta\nu\over
		  \nu_{1720}\ \nu_{1612}\right)_\circ    
  \label{eqn:linediff}
\end{equation}
The conjugate OH lines in PKS 1413+135 provide an ideal laboratory for
$\alpha$ measurements because the lines are guaranteed to arise from the same
molecular gas.  From the OH lines alone, we obtain 
$\Delta\alpha/\alpha_\circ = (0.5\pm1.3)\times10^{-5}$, consistent with
no change in $\alpha$ (Table \ref{table:alpha}).  
The precision of this OH-only measurement could be
significantly improved by comparing the 18 cm OH transitions to higher 
frequency OH transitions at 4 and 6 GHz.

\begin{figure*}
\epsscale{1.80}
\plotone{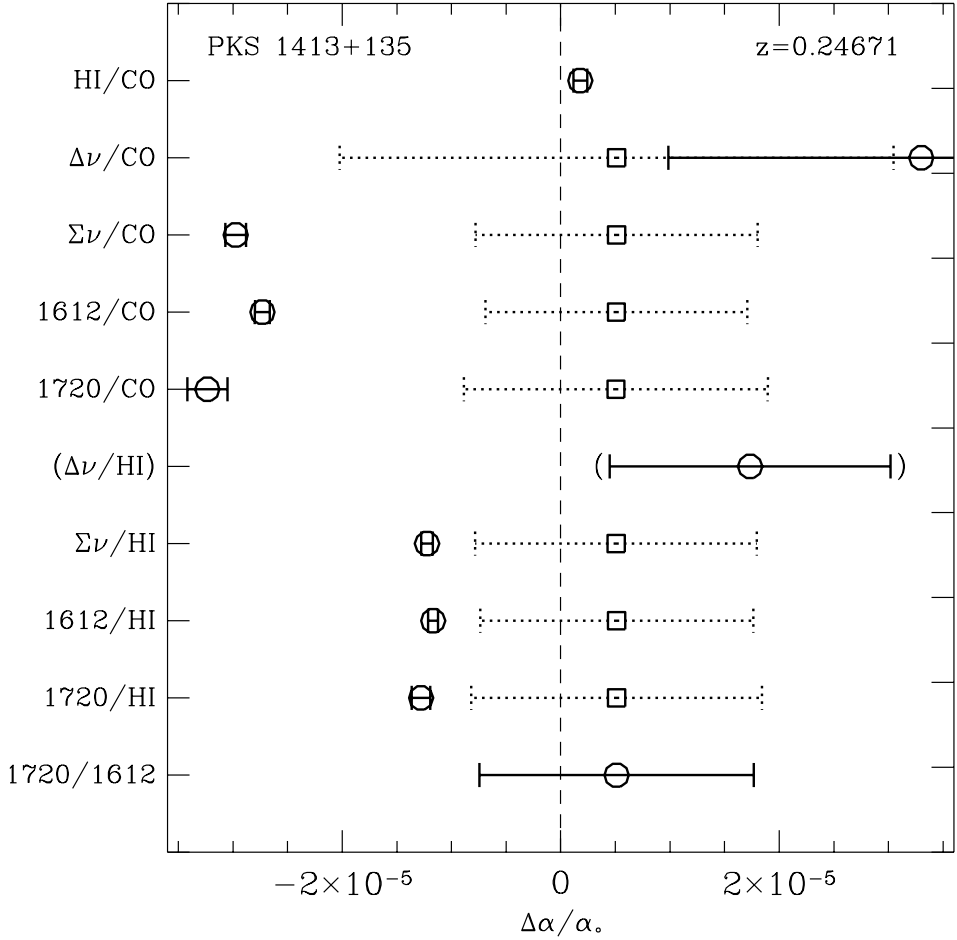}
\caption{Changes in the fine structure constant measured 
from pairs of OH, \ion{H}{1}, and CO lines at $z=0.24671$ toward PKS 1413+135.
Circles with solid error bars are apparent changes measured from 
redshift pairs, and the squares with 
dotted error bars are 
$\Delta\alpha/\alpha_\circ$ values corrected for systematic
offsets between species.
Note that from four lines, there are only three independent 
apparent $\Delta\alpha/\alpha_\circ$ values, and only two independent 
systematics-corrected values.
The $\Delta\nu/$\ion{H}{1} quantity is not a true measure of
$\Delta\alpha/\alpha_\circ$; it is a measure of the $\alpha$-independent
systematic offset between the OH and HI redshifts, translated into
a $\Delta\alpha/\alpha_\circ$ measure to indicate the magnitude of the
correction applied to the apparent OH-\ion{H}{1} values.
The OH and \ion{H}{1} lines
were observed in a barycentric frame, but the CO reference frame is 
heliocentric; the difference between frames is of order 10 m s$^{-1}$, 
which is an order of magnitude less than the uncertainty in the \ion{H}{1} and
CO line centroids.  The CO redshift refers to the strongest CO 
$J=0\rightarrow1$ line
observed by \citet{wik97}.\label{fig:alpha}}
\end{figure*}

The OH redshifts can be compared to molecular line or \ion{H}{1} redshifts for
additional independent measurements of $\alpha$, and one can employ the
pure $\Lambda$-doubled and pure hyperfine OH quantities to constrain 
systematic velocity offsets and measurement errors between species.  In 
particular, $z_{HI}-z_{\Delta\nu} = 0$  in the absence 
of systematic offsets between species.  Any nonzero value obtained from this
expression thus quantifies the systematic errors in other \ion{H}{1}-OH determinations
of $\alpha$.  Table \ref{table:alpha} lists and Figure \ref{fig:alpha}
plots apparent $\Delta\alpha/\alpha_\circ$ 
values from pairs of OH, \ion{H}{1}, and CO lines 
(note that from four lines, there are only three independent values).
These values are computed from OH satellite line 
analogues to the main line expressions presented in \citet{dar03}:
\begin{eqnarray}
   {z_{HI}- z_{1612} \over 1+z_{1612}}
		&\simeq& -1.8\,{\Delta\alpha\over\alpha_\circ}
		\left({\Sigma\nu\over \nu_{1612}}\right)_\circ\\
   {z_{HI}- z_{1720} \over 1+z_{1720}}
		&\simeq& -1.8\,{\Delta\alpha\over\alpha_\circ}
		\left({\Sigma\nu\over \nu_{1720}}\right)_\circ\\
   {z_{CO}- z_{1612} \over 1+z_{1612}}
		&\simeq& -{\Delta\alpha\over\alpha_\circ}
	\left({0.8\,\Sigma\nu+\Delta\nu\over\nu_{1612}}\right)_\circ\\
   {z_{CO}- z_{1720} \over 1+z_{1720}}
		&\simeq& -{\Delta\alpha\over\alpha_\circ}
		\left({0.8\,\Sigma\nu-\Delta\nu\over \nu_{1720}}\right)_\circ
\end{eqnarray}
The expressions for comparisons of CO and \ion{H}{1} to $\Sigma\nu$ and 
$\Delta\nu$ are identical to those presented in \citet{dar03}.

OH-\ion{H}{1} and OH-CO line pairs show highly significant offsets in the 
{\it apparent} value of $\alpha$ at $z=0.24671$ 
($\Delta\alpha/\alpha_\circ = (-1.223\pm0.054)\times10^{-5}$ and 
 $\Delta\alpha/\alpha_\circ = (-2.98\pm0.10)\times10^{-5}$ respectively), 
but are these offsets 
due to a changing $\alpha$ or are they due to systematic errors and/or
velocity offsets between species?  We are confident that the simultaneous
observations of OH and \ion{H}{1} conducted at the GBT produce reliable 
relative
offsets between lines and we can thus exclude many systematic effects related
to the atmosphere, telescope, doppler tracking, backend, and data reduction
pathways.  To assess the possibility that the offset is a physical velocity
difference between the \ion{H}{1} and OH gas, we employ the pure hyperfine  
$z_{HI}-z_{\Delta\nu}$ estimate of the systematics:
$\Delta z_{sys}(OH-HI) = (7.78\pm5.77)\times10^{-5}$, which is 
$18.7\pm13.9$ km s$^{-1}$ in the rest frame of PKS 1413+135.
This quantity, translated
into a $\Delta\alpha/\alpha_\circ$ offset of $(+1.73\pm1.29)\times10^{-5}$, 
shows that the systematic
offsets, while uncertain in magnitude, can account for {\it all} of the
apparent change in $\alpha$ in OH-\ion{H}{1} line pairs.  
A similar argument applies
to the OH-CO line pairs, but  since OH lines cannot form a quantity 
which depends on $\alpha^2$, the $\Sigma\nu$-CO and $\Delta\nu$-CO 
quantities bracket the true value of $\Delta\alpha/\alpha_\circ$ and should
produce consistent values in the absence of systematic effects.  Since they
are not consistent in PKS 1413+135, we can solve for
the systematic velocity offsets between OH and CO (see Appendix 
\ref{appendix}).  We obtain
$\Delta z_{sys}(OH-CO) = (6.95\pm2.57)\times10^{-5}$, which is
$16.7\pm6.2$ km s$^{-1}$ in the rest frame.
Hence, the OH-CO
systematic offsets can account for the apparent change in $\alpha$ in
OH-CO line pairs as well.   The inter-species systematic corrections have
been applied to the appropriate line pairs and are listed in Table 
\ref{table:alpha} and plotted in Figure \ref{fig:alpha}.
Finally, the \ion{H}{1}-CO pair indicates a 2.9$\sigma$
deviation from $\Delta\alpha=0$ that the OH observations suggest could
be completely attributed to systematic effects, as suggested by \citet{car00}.

Is is no coincidence that all corrected points in Figure \ref{fig:alpha} 
have the same value and nearly the same uncertainties.  The correction for
systematic offsets between species relies on the measurement of $\Delta\nu$
to quantify the systematics, but this difference in frequencies is precisely
the quantity used to derive $\Delta\alpha/\alpha_\circ$ from the 1720-1612
MHz line pair of OH.  In removing the systematic effects, we have 
used the OH redshift as a fiducial, so the corrected points are only
as precise as the OH zero point.  Correction for $\alpha$-independent 
effects is equivalent to requiring all measurements to produce the same value
of $\Delta\alpha/\alpha_\circ$, which is exactly what the correction procedure
has produced.  The systematics corrections thus reduce the number of
independent $\Delta\alpha/\alpha_\circ$ values to just two.
The correction for systematic effects between species 
reveals that --- contrary to the small statistical error bars on apparent 
measurements between species --- the most precise measurement involving
OH lines is the OH-only 
measurement obtained from the OH satellite lines.

\section{Conclusions}

Our observations of
the 18 cm OH lines support a scenario of radiatively pumped stimulated 
absorption and emission with pumping dominated by the intraladder 119 
$\mu$m line of OH, 
suggesting a column density $N(OH) \simeq 10^{15}$--$10^{16}$ cm$^{-2}$.
The OH alone, in which conjugate line profiles
guarantee that both lines originate in the same molecular gas, 
provides a weak constraint on the evolution of the fine structure constant
of $\Delta\alpha/\alpha_\circ = (+0.51 \pm 1.26) \times 10^{-5}$. This 
result illustrates the need for higher frequency OH line detections to provide
a larger lever arm on $\Delta\alpha$, but it also illustrates the power of the
OH molecule to provide precise measurements of the cosmic evolution 
of $\alpha$ that include quantitative constraints on systematic errors.  
We obtain a highly significant ($\sim25\sigma$)
velocity offset between the OH and \ion{H}{1} and the OH and CO lines 
which is
entirely attributable to systematic velocity offsets between species.  
The OH-HI offset is likely to be a physical velocity offset (perhaps due to 
different physical locations in the extended absorber), and the OH-CO
offset may be a combination of physical and measurement-based systematics.  
We obtain a marginally significant ($2.9\sigma$) HI-CO offset which the OH
observations indicate is likely due to systematic effects,
as anticipated by \citet{car00}.

Detection of conjugate OH lines at higher redshifts will provide new
constraints on the cosmic evolution of physical constants.  While the
detection of new centimeter and millimeter molecular absorption systems 
has proved to be a difficult 
and thus far fruitless task, recent and future improvements in 
frequency coverage, instantaneous
bandwidth, sensitivity, and RFI mitigation should facilitate 
new discoveries in the near future.  There is room for optimism for 
detecting conjugate OH lines at high redshift because, like absorption
lines and weak maser lines, the strength of conjugate OH lines 
depends only on the flux density of the background radio continuum source
and not on the redshift.  Also, as \citet{fra98} point out, the 
density, temperature, and molecular column density conditions required 
to form conjugate OH lines are common in the centers of active galaxies, so
conjugate OH lines may be most efficiently identified in searches for
intrinsic absorption in radio loud active galaxies rather than in blind
searches for intervening absorption line systems.



\acknowledgments

We are grateful to the hardworking staff at NRAO Green Bank 
for observing and data reduction support, especially Karen O'Neil, 
Bob Garwood, and Carl Bignell.  We also thank the anonymous referee for
constructive and insightful comments.






\appendix

\section{Quantifying Systematic Offsets Between Species}
\label{appendix}

Redshift offsets between species with the same $\alpha$ dependence will 
to first order quantify the net systematic effects, be they instrumental, 
propagation effects, or physical velocity differences between species.  
Hence, in the case of \ion{H}{1} and the (to first order) pure hyperfine
quantity $\Delta\nu$ derived from the frequency differences between 
pairs of main or satellite OH lines, we
expect redshift differences to be a measure of systematics rather than 
changes in fundamental constants:  $z_{HI}-z_{\Delta\nu} = \Delta z_{sys}$.
The derived $\Delta z_{sys}$ should then be applied as a correction to 
the $\Delta\alpha/\alpha_\circ$ values obtained from \ion{H}{1}-OH line
pairs.  

Systematic corrections can also be measured and applied in cases where 
there is no convenient cancellation of $\alpha$ terms; only different 
$\alpha$ dependencies are required.  For example, comparison of the 
sum and difference, $\Sigma\nu$ and $\Delta\nu$, of an OH 
main or satellite line pair to a transition in species 
$X$ with dependence $\alpha^\gamma$ will give
\begin{eqnarray}
 z_X - z_{\Sigma\nu} &=& 
	\beta_\Sigma(1+z_{\Sigma\nu}){\Delta\alpha\over\alpha_\circ} + 
	\Delta z_{sys} \\
 z_X - z_{\Delta\nu} &=& 
	\beta_\Delta(1+z_{\Delta\nu}){\Delta\alpha\over\alpha_\circ} + 
	\Delta z_{sys} 
\end{eqnarray}
where $\beta_\Delta=4-\gamma$, $\beta_\Sigma=\eta-\gamma$ with
$\eta = 0.4$ for $^2\Pi_{3/2}$ and $\eta=5$ for $^2\Pi_{1/2}$ states, and 
$\Delta z_{sys}$ represents an average $\alpha$-independent offset 
between species.  The requirement of a consistent measurement of 
$\Delta\alpha/\alpha_\circ$ from both line comparisons provides a 
solution for $\Delta z_{sys}$:
\begin{equation}
  \Delta z_{sys} = z_X + {z_{\Sigma\nu}\,\beta_\Delta - 
	z_{\Delta\nu}\,\beta_\Sigma \over \beta_\Sigma-\beta_\Delta}\:.
\end{equation}
In the case where $\beta_\Sigma=\beta_\Delta$, there is no leverage on 
$\Delta z_{sys}$ and systematics cannot be quantified.  In the case
where $\beta_\Sigma=0$ or $\beta_\Delta=0$ (which is the case for \ion{H}{1}),
$\Delta z_{sys}$ can be 
measured directly from the $\alpha$-canceled line pair.




\clearpage



\clearpage

\clearpage



\clearpage

\begin{deluxetable}{ccccccc}
\tablewidth{0pt}
\tabletypesize{\footnotesize}
\tablecaption{Observed and Fit Properties of OH and \ion{H}{1} Lines Toward
PKS 1413+135
 \label{table:fits}}
\tablehead{\colhead{Line} & \multicolumn{5}{c}{Gaussian Fits}
&\colhead{$\int\tau\ dV$}\\ \cline{2-6}\relax\\[-1ex]
\colhead{}&\colhead{$\nu_c$}&\colhead{$z_c$}
&\colhead{$S_c$}&\colhead{$\tau_c$}&
\colhead{$\Delta V$} \\
\colhead{}&\colhead{MHz}&\colhead{}    &\colhead{mJy}  &\colhead{}&
\colhead{km s$^{-1}$}&\colhead{km s$^{-1}$}
}
\startdata
OH 1612 & $1293.2502\pm0.0015$ & $0.2466505\pm0.0000014$ & 
	$-11.67\pm0.93$ & $0.0098\pm0.0008$ & $8.7\pm0.8$ & 0.091 \\
OH 1665\tablenotemark{a} & \nodata & \nodata & (1.3) & (0.0011) & \nodata & 
	\nodata\\
OH 1667\tablenotemark{a} & \nodata & \nodata & (1.1) & (0.0009) & \nodata & 
	\nodata\\
OH 1720 & $1380.1238\pm0.0037$ & $0.2466490\pm0.0000034$ & 
	$+11.7\pm2.0$ & $-0.0097\pm0.0017$ & $10.1\pm1.9$& $-0.108$\phs \\
\ion{H}{1} 1\tablenotemark{b}& $1139.3282\pm0.0014$ & $0.2467046\pm0.0000015$&
	$-594\pm20$ & $0.68\pm0.03$ & $12.6\pm1.8$ & 13.4\tablenotemark{c} \\
\ion{H}{1} 2    & $1139.3660\pm0.0026$ &  $0.2466633\pm0.0000028$ &
	$-165\pm52$ & $0.15\pm0.05$ & $8.1\pm1.2$&  \\
\ion{H}{1} 3    & $1139.2886\pm0.0193$ & $0.246748\pm0.000021$ & 
	$-76\pm55$ & $0.07\pm0.05$ & $12.3\pm5.4$ &  \\
\ion{H}{1} 4    & $1139.2775\pm0.0112$ & $0.246760\pm0.000012$ & 
	$-36\pm5$ & $0.030\pm0.004$ & $65.8\pm6.8$ &  \\
\enddata
\tablenotetext{a}{The 1665 and 1667 MHz lines were not detected.  The 
listed flux densities and optical depths are rms noise values in boxcar
smoothed spectra with 4.0 km s$^{-1}$ resolution similar to the resolution
of the OH 1667 MHz observations of PKS 1413+135 by \citet{kan02}.}
\tablenotetext{b}{This is the dominant \ion{H}{1} line, and it
is used to determine
the \ion{H}{1} redshift.  Note that the blending with other lines makes the peak
frequency of the main \ion{H}{1} line more uncertain than would typically be 
expected for such a high signal-to-noise spectrum.}
\tablenotetext{c}{This integrated optical depth includes all observed \ion{H}{1}
absorption.}
\end{deluxetable}

\begin{deluxetable}{cccccc}
\tablewidth{0pt}
\tabletypesize{\footnotesize}
\tablecaption{Fine Structure Constant Evolution from OH, \ion{H}{1}, and CO Lines}
\tablehead{& & & & 
\multicolumn{2}{c}{$\Delta\alpha/\alpha_\circ$} \\
\cline{5-6}
\colhead{Line 1} & \colhead{Line 2} 
& \colhead{$z_1$}& \colhead{$z_2$}      &Apparent & Corrected \\
& & & & $[\times10^{-5}]$ & $[\times10^{-5}]$
}
\startdata
OH 1720	& OH 1612 & $0.2466490\pm0.0000034$ & 
	$0.2466505\pm0.0000014$ & $+0.51\phd\pm1.26\phd$ & \nodata \\
OH 1720	& \ion{H}{1} 1    & $0.2466490\pm0.0000034$ & 
	$0.2467046\pm0.0000015$ & $-1.278\pm0.085$ &
	$+0.51\phd\pm1.33\phd$  \\
OH 1612	& \ion{H}{1} 1    & $0.2466505\pm0.0000014$ & 
	$0.2467046\pm0.0000015$ & $-1.167\pm0.045$ &
	$+0.51\phd\pm1.25\phd$  \\
OH $\Sigma\nu$& \ion{H}{1} 1& $0.2466498\pm0.0000019$ & 
	$0.2467046\pm0.0000015$ & $-1.223\pm0.054$ &
	$+0.51\phd\pm1.29\phd$  \\
OH $\Delta\nu$& \ion{H}{1} 1& $0.246627\pm0.000058$ & 
	$0.2467046\pm0.0000015$ & $+1.73\phd\pm1.29$\tablenotemark{a}
	& \nodata \\
OH 1720	& CO ($0\rightarrow1$)& $0.2466490\pm0.0000034$ & 
	$0.2467091\pm0.0000003$ & $-3.23\phd\pm0.18\phd$ &
	$+0.51\phd\pm1.39\phd$ \\
OH 1612	& CO ($0\rightarrow1$)& $0.2466505\pm0.0000014$ & 
	$0.2467091\pm0.0000003$ & $-2.73\phd\pm0.07\phd$ &
	$+0.51\phd\pm1.20\phd$ \\
OH $\Sigma\nu$& CO ($0\rightarrow1$)& $0.2466498\pm0.0000019$ & 
	$0.2467091\pm0.0000003$ & $-2.98\phd\pm0.10\phd$ &
	$+0.51\phd\pm1.29\phd$ \\
OH $\Delta\nu$& CO ($0\rightarrow1$)& $0.246627\pm0.000058$ & 
	$0.2467091\pm0.0000003$ & $+3.30\phd\pm2.32\phd$ & 
	$+0.51\phd\pm2.53\phd$ \\
\ion{H}{1} 1 & CO ($0\rightarrow1$)& $0.2467046\pm0.0000015$ & 
	$0.2467091\pm0.0000003$ & $+0.179\pm0.063$  & \nodata \\
\enddata
\tablenotetext{a}{This is not a true measure of $\Delta\alpha/\alpha_\circ$;
it is a measure of the systematic offset between the OH and \ion{H}{1} lines translated
into a $\Delta\alpha/\alpha_\circ$ offset.}
\tablerefs{OH, \ion{H}{1} redshifts:  this work; 
CO ($0\rightarrow1$) redshift:  \citet{wik97}.}
\label{table:alpha}
\end{deluxetable}




\end{document}